\documentclass[aps,prl,reprint]{revtex4-2}

\usepackage{amsmath, amsthm, bm}
\usepackage{amssymb}
\usepackage{mathrsfs}
\usepackage{amsfonts}
\usepackage{graphicx}
\usepackage{float}
\usepackage[caption=false]{subfig}
\usepackage[dvipsnames]{xcolor}
\usepackage{siunitx}
\usepackage{bbold}
\usepackage[version=3]{mhchem}
\usepackage[acronym]{glossaries}
\usepackage{array}
\usepackage{blkarray}
\usepackage{tikz}
\usetikzlibrary{arrows}
\usepackage{xfrac}

\usepackage[section]{placeins}
\usepackage{indentfirst}

\usepackage{xcolor}

\DeclareSIUnit\au{\text {au}}
\DeclareSIUnit\angstrom{\text {Å}}

\newacronym{vsc}{VSC}{vibrational-strong coupling}
\newacronym{esc}{ESC}{electronic-strong coupling}
\newacronym{lp}{LP}{lower polariton}
\newacronym{up}{UP}{upper polariton}
\newacronym{tc}{TC}{Tavis-Cummings}
\newacronym{fc}{FC}{Franck-Condon}

\newcommand{\da}{^\dagger}

\newcommand{\br}[1]{\langle #1 \vert}
\newcommand{\ke}[1]{\vert #1 \rangle}

\newcommand{\out}[2]{\vert #1 \rangle \langle #2 \vert }
\newcommand{\ev}[1]{\langle #1 \rangle}

\newcommand{\cm}[2]{\left[#1,#2\right]}

\newcommand{\destroy}{\hat a}
\newcommand{\create}{\hat a^\dagger}

\newcommand{\sigmap}{\hat\sigma^+}
\newcommand{\sigmam}{\hat\sigma^-}
\newcommand{\sigmapm}{\hat\sigma^\pm}

\newcommand{\Sigmap}{\hat S^+}
\newcommand{\Sigmam}{\hat S^-}
\newcommand{\Sigmapm}{\hat S^\pm}

\newcommand{\hamilt}{\hat{\mathcal H}}
\newcommand{\ith}{$i$-th~}

\newcommand{\tr}[1]{\text{Tr}\left[#1\right]}

\newcommand{\Nexc}{N_\text{x}}
\newcommand{\NexcTC}{N_\text{x,TC}}
\newcommand{\hNexc}{\hat{N}_\text{x}}
\newcommand{\eV}{\text{eV}}
\newcommand{\fs}{\text{fs}}

\begin{document}
\title{The role of dark polariton states for electronic strong coupling in molecules}

\author{Lucas Borges}
\affiliation{Department of Physics, Stockholm University, AlbaNova University Center, SE-106 91 Stockholm, Sweden}

\author{Thomas Schnappinger}
\affiliation{Department of Physics, Stockholm University, AlbaNova University Center, SE-106 91 Stockholm, Sweden}

\author{Markus Kowalewski}
\email{markus.kowalewski@fysik.su.se}
\affiliation{Department of Physics, Stockholm University, AlbaNova University Center, SE-106 91 Stockholm, Sweden}

\date{\today}%

\begin{abstract}
Polaritonic chemistry investigates the possible modification of chemical and photochemical reactions by means of strong light-matter coupling in optical cavities, as demonstrated in numerous experiments over the last few years.
These experiments are typically interpreted in terms of the Jaynes-Cummings or Tavis-Cummings models under the assumption that the molecular ensemble is only excited by a single photon.
In such a model, two polariton states compete with an overwhelming number of dark states, inhibiting polaritonic reactions entropically.
We analyze the higher excitation manifolds of the Tavis-Cummings model along with a three-level system that resembles photochemical reactions. 
We demonstrate that allowing for more than a single excitation makes the reaction of the involved polaritons entropically more favorable.
\end{abstract}

\maketitle

Polaritons are hybrid light-matter states that can form when matter transitions are strongly coupled to the electromagnetic field in an optical cavity~\cite{Baranov2018-hm,Herrera2020-bg,Basov2021-hr,Garcia-Vidal2021-qe,Sanchez-Barquilla2022-dq,Li2022-gi}. 
These states occur when the light-matter interaction rate is faster than the individual photon and matter decay rates in the system.
Molecular polaritons have become an emerging area of research at the interface of quantum optics, chemistry, and materials science~\cite{Ruggenthaler2018-ew,Fregoni2022-op,Dunkelberger2022-oh,Mandal2023-ob,Bhuyan2023-se,Hirai2023-qm,Gu2023-et,Xiang2024-nn}. 
Exciton polaritons, which combine electronic excitations with confined light modes, can alter the properties of excitonic systems, influencing processes such as electron transport~\cite{Orgiu2015-fd,Eizner2019,Polak2020-ac,Sokolovskii2023-zk,Balasubrahmaniyam2023-nc,Wallner2024-qg,Koessler2025-pz}, light harvesting~\cite{Esteso2021-ox,Wu2022-xs}, or energy transfer~\cite{Coles2014-yl,Zhong2016-ao,Zhang2024-kt}.
When confined light modes are coupled to molecular vibrations, changes in molecular properties~\cite{Fukushima2022-ox} and chemical reactivity~\cite{Thomas2019-ve,Ahn2023-qk} have been reported from experiments.
Modified and extended versions of the Rabi and Dicke model~\cite{Rabi1937-jq,Dicke1954-sq} or the Jaynes-Cummings and \gls{tc} model~\cite{Jaynes1963-re,Tavis1967-op} are often used to interpret the formation of hybrid polaritonic states and the outcome of polaritonic experiments.
Despite considerable efforts, the fundamental theoretical understanding of these experiments is still incomplete.

Beyond these quantum-optical models, more recently ab-initio methods such as quantum electrodynamics density functional theory \cite{Ruggenthaler2014-it} and coupled cluster theory~\cite{Haugland2020-xh} have been developed.
However, in contrast to these methods, \gls{tc} models offer a particularly efficient way to simulate large numbers of molecules~\cite{Tichauer2021-mk,Angulo_2021,Cui2022-po,Zhou2023-zo,Sokolovskii2023-zk,Perez-Sanchez2023-mc}.
As shown in the literature~\cite{Groenhof2019-nz,Davidsson2020-bs,Davidsson2023-xa,Gudem2021-um,Borges2024-pn}, \gls{tc} models can also be extended to include aspects such as nuclear degrees of freedom and permanent dipole moments relevant for molecular systems.
Another advantage of models based on the \gls{tc} Hamiltonian is that they allow a straightforward separation of the states of the coupled system into manifolds with an equal number of excitations in the molecular and photonic parts.
Much of the research applying these models has focused on the first excitation manifold, where the formation of two polaritonic states \gls{lp} and \gls{up}, as well as dark states, in the case of more than one molecule is well understood. 
These dark states are decoupled from the cavity field and do not participate in the light-matter interaction but are still important for the dynamics of the system, especially in open quantum systems~\cite{Ulusoy2020-qk,Antoniou2020-lp,Davidsson2020-bs,Davidsson2023-xa, Ballestero2016,Groenhof2019-nz}. 
In the probably more realistic scenario and allowing more excitations in the coupled system~\cite{DelPo,Campos-Gonzalez-Angulo2022-rz,Siltanen2025}, additional types of hybrid state arise.
The excitation of polaritons and dark states associated with lower excitation manifolds gives rise to different types of hybrid state, called multi polaritons and dark polaritons in the literature~\cite{DelPo,Campos-Gonzalez-Angulo2022-rz}. 
These new additional hybrid states lead to more complex eigenvalue patterns in the higher excitation manifolds.

In this work, we investigate the influence of the eigenvalue structure of higher excitation manifolds of the \gls{tc} model on reactions under the influence of electronic strong coupling.
We extend the \gls{tc} model to a three-level system that allows us to model photochemical reactions, such as singlet fission or triplet-triplet annihilation, which can be influenced by strong light-matter coupling~\cite{Orgiu2015-fd,Eizner2019,Polak2020-ac,Ye2021-sc,Sokolovskii2023-zk,Balasubrahmaniyam2023-nc,Dutta2024-jm,Wallner2024-qg}.
The manuscript is structured as follows. In the first part, the Hamiltonian for the standard \gls{tc} model and extended \gls{tc} model with an additional excited molecular state is introduced.
To set the stage, we discuss the eigenvalue structure of the \gls{tc} Hamiltonian and its extended version for higher excitation manifolds.
In the final part, we present dissipative dynamics for the model systems and discuss the mechanisms and difference with the first excitation manifold.

All simulations in this study are based on the \gls{tc} Hamiltonian~\cite{Tavis1967-op} for $N$ identical two-level molecules. 
Here, we assume that the electronic states of a molecule can be described by two states $\ke{g}$ and $\ke{e}$, which are coupled to a single mode of an optical cavity.
In the following, we use atomic units ($\hbar = 4\pi \varepsilon_0 = m_e = 1$).
The \gls{tc} Hamiltonian reads:
\begin{equation}
	\hamilt_{TC} =  \omega_{eg}\sum_{i=1}^N\ke{e_i}\br{e_i}+
	  \omega_c\create \destroy + g_c \left(\create\Sigmam +\destroy\Sigmap \right)\,, \label{eq:TCHamiltonian}
\end{equation}
where $\omega_{eg} = \omega_e-\omega_g$ is the energy difference between the two molecular energy levels $\ke{e}$ and $\ke{g}$, and $\omega_c$ is the frequency of the cavity mode.
Here, $\hat{a}^\dagger$ and $\hat{a}$ are the bosonic creation and annihilation operators for the cavity mode.
The coupling of the molecular excitation and the photon mode in the dipole approximation \cite{Schleich2001} is described by the cavity coupling strength $g_c = \mu_{ge}\sqrt{4\pi\omega_c/V_c}$,
where $V_c$ is the quantization volume of the cavity and $\mu_{ge}$ is the transition dipole moment between the states $\ke{g}$ and $\ke{e}$.
The first term in Eq.~\ref{eq:TCHamiltonian} describes the molecular excitations, the second term describes the excitation of the cavity mode, and the third term describes the coupling of molecular excitations to the cavity mode under the assumption of the rotating wave approximation~\cite{Schleich2001}.
The molecular excitation and de-excitation operators $\Sigmapm$ are given by 
\begin{equation}	
	\Sigmapm = \sum_{i=1}^{N} \sigmapm_i\,
\end{equation}
where $\sigmap_i = (\sigmam_i)^\dagger = \out{e_i}{g_i} $ is the local Pauli operator exciting the \ith molecule from state $\ke{g}$ to $\ke{e}$.
The operator representing the total number of excitations is given by 
\begin{equation}
    \hat N_{x,TC} = \sum_{i=1}^{N} \sigmap_i\sigmam_i + \create\destroy ,
\end{equation}
where the first term yields the total number of excited molecules, and the second term, $\hat n = \hat a\da\hat  a$, yields the number of photons in the cavity mode. 
$\hat{N}_{x,TC}$ commutes with the Hamiltonian in Eq.~\eqref{eq:TCHamiltonian}, and its eigenstates can be grouped into sets of states that share the same number of excitations, which we refer to as excitation manifolds. 
Furthermore, we define the $\hat S^2$ operator
\begin{align}
    \hat S^2 \ke{\Psi} = S(S+1)\ke{\Psi}\,,
\end{align}
via its eigenvalues, which will be used to classify groups of eigenstates and is a conserved quantity in the absence of dissipation~\cite{Dicke1954-sq}.
The cooperation number $2S=0\dots N$ can be interpreted as the number of molecules that contribute to a collectively coupled state.

The eigenstates of the first excitation manifold with respect to Eq.~\ref{eq:TCHamiltonian}, are the well-known \gls{lp} and \gls{up} states and a set of $N-1$ dark states for which the molecules are decoupled from the cavity mode.
The observable effective Rabi-splitting scales with the number of particles actively coupled to the cavity mode,
and for $\NexcTC \ll N$ and $N \gg 1$ we can write:
\begin{equation}
\Omega_R \approx \sqrt{ 8g_c^2 S+ \left(\omega_{eg}-\omega_{c}\right)^2 }\,.
\label{eq:Rabi}
\end{equation}
Note, that the effective splitting now depends on $2S$ which coincides with $N$ for the maximally allowed value of $S$.

\begin{figure}
\includegraphics[width=7cm]{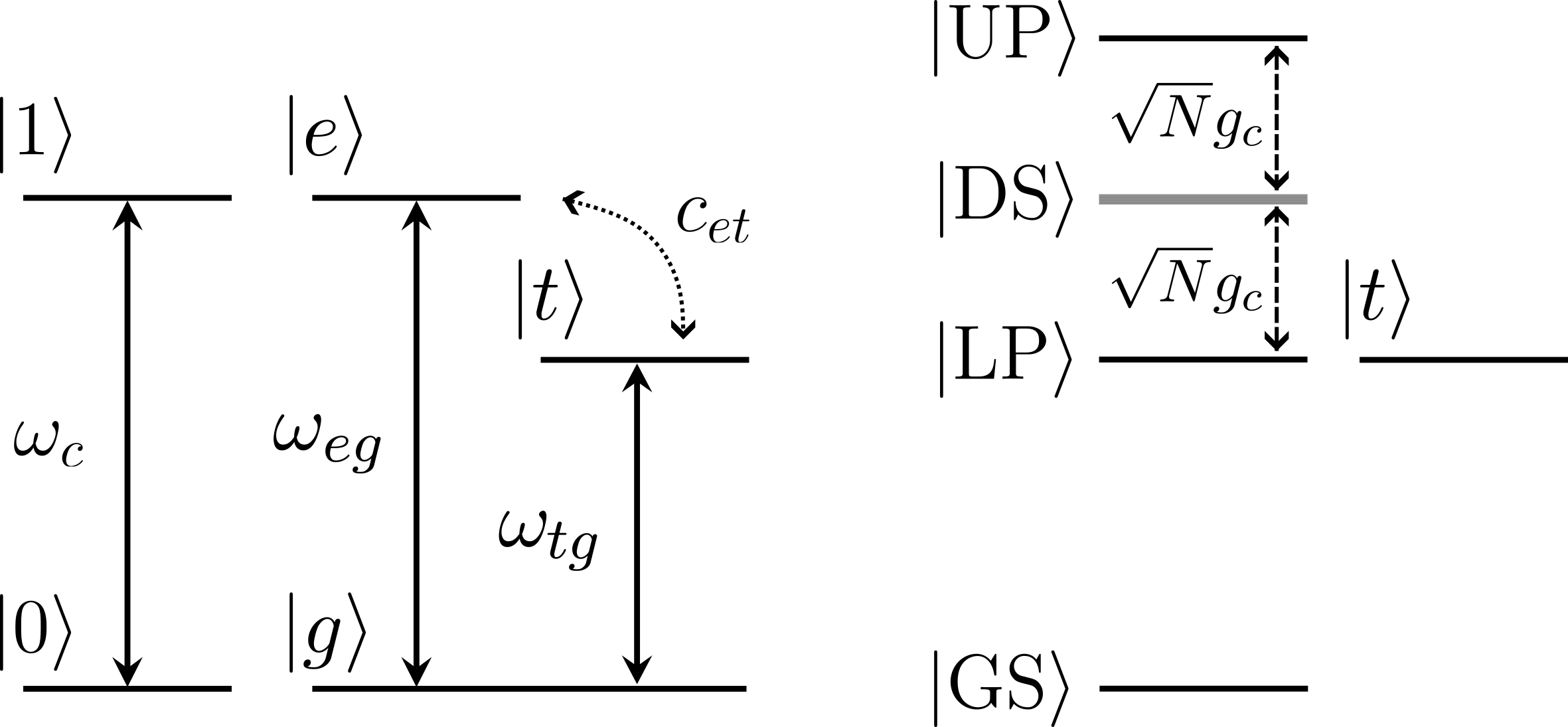}
\caption{Three-level system scheme. Left panel: bare levels of the cavity and molecular systems for $N=1$. Right panel: eigenstates for $N>1$ and $\Nexc = 1$. The cavity coupling between the excited state $\ke{e}$ and the ground state $\ke{g}$ leads to the formation of polariton states (\gls{lp} and \gls{up}), as well as dark states, which are coupled to a third state $\ke{t}$.}\label{fig:tls}
\end{figure}

Next, we introduce a third molecular state $\ke{t}$, which is coupled to the excited state $\ke{e}$ through a constant coupling $c_{et}$.
Such a coupling may be caused, for example, by a non-adiabatic process or internal conversion between a singlet and a triplet state.
We assume that $\ke{t}$ is optically dark, i.e., has no transition dipole moment with respect to states $\ke{g}$ or $\ke{e}$, and therefore does not couple with the cavity mode.
Figure~\ref{fig:tls} shows the corresponding level diagrams for the uncoupled states and the eigenstates in the case of resonant coupling to a cavity.
The bare energies of $\ke{t}$ and $\ke{e}$ are detuned in the given example, making the population transfer ineffective despite the coupling.
However, the lower polariton state can be brought into resonance with $\ke{t}$, making the hybridization of $\ke{t}$ and $\ke{e}$ more efficient.

\begin{figure*}
\centering
\includegraphics[width=15cm]{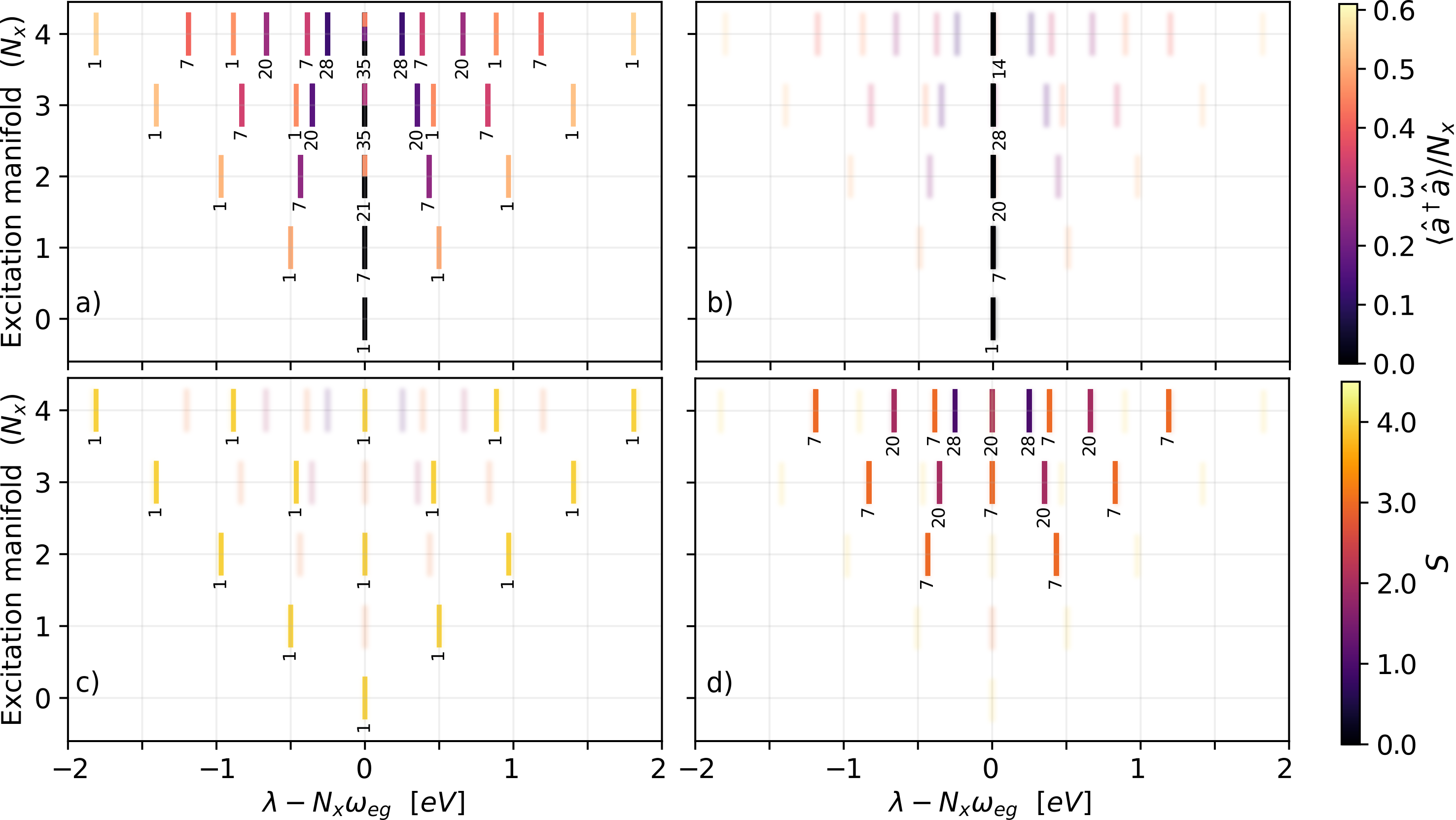}
\caption{Energy diagrams for the eigenvalues ($\lambda$) of $N=8$ two-level molecules resonantly coupled to a single cavity mode. The eigenenergies are shown as excitation number versus relative energy shift.
The color code in panel (a) and (b) indicates the normalized photonic character ($\ev{\hat a^\dagger \hat a}/\NexcTC$) of each eigenstate group. We decompose these eigenstate groups into (b) dark states, (c) multi polaritons ($S=N/2$), and (d) dark polaritons ($S<N/2$ and $\ev{\hat n}>0$). The inset numbers represent the number of degenerate states per eigenvalue. The color code in (c) and (d) indicates the cooperation number $S$. The parameters are: $g_c=0.5\eV/\sqrt N, \omega_{eg}= \omega_c = 4.3\eV$.}
\label{fig:ladder_2LS_N8}
\end{figure*}

The Hamiltonian from Eq.~\eqref{eq:TCHamiltonian} is extended to a molecular three level system, which then reads:
\begin{align}
    \hamilt = \hamilt_{TC} + \omega_{tg}\sum_{i=1}^{N} \out{t_i}{t_i} +  c_{et}\left(\Sigmap_t + \Sigmam_t \right), 
\label{eq:3LS_Hamilt}
\end{align}
where $\omega_{tg} = \omega_{t}-\omega_{g}$ is the energy difference between the states $\ke{g}$ and $\ke{t}$.
The operators 
\begin{equation}	
	 \Sigmap_t = (\Sigmam_t)^\dagger = \sum_{i=1}^{N} \out{e_i}{t_i},
\end{equation}
describe the transitions between $\ke{t}$ and $\ke{e}$ in the $i$th molecule.
The number operator for the extended three-level system can then be written as $\hNexc =  \hat N_\text{x,TC} + \hat N_t$,
where $\hat N_t = \sum_{i=1}^{N} \out{t_i}{t_i}$. 
Since $\hNexc$ commutes with $\hamilt$, $\hNexc$ can be used to group the eigenstates by their excitation number.
Both levels $\ke{e}$ and $\ke{t}$ are considered excited states, and thus the total number of states for a given $\Nexc \le N$  is given by $\sum_{m=0}^{\Nexc} 2^m \binom{N}{m}$.

The Fabry-P\'erot cavities that are typically used in polaritonic experiments~\cite{Mony18JPCC} have Q factors on the order of 100.
Thus, photon decay from the cavity mode is an important process driving photochemical reactions described by the aforementioned three-level system.
In addition, the dissipative channels that are opened by cavity decay and spontaneous emission from $\ke{e}$ connect subspaces of different cooperation numbers $S$ that are otherwise disconnected \cite{Davidsson2023-xa,Quesada2012-vd} in a pure Schr\"odinger picture.

To describe the dynamics of the open-system and the density matrix $\rho$, we use a Lindblad master equation:
\begin{align}\label{eq:master_equation}
    \dot \rho(t) =& -i\cm{\hamilt(t)}{\rho(t)} \\ &-
    \frac{\kappa}{2}\left(\create\destroy\rho(t) + \rho(t)\create\destroy - 2\destroy\rho(t)\create \right) \nonumber
    \\&+ \sum_{i=1}^N\frac{\Gamma}{2}\left(\rho(t)\out{e_i}{e_i} + \out{e_i}{e_i}\rho(t) - 2\sigmam_i\rho(t)\sigmap_i \right)\nonumber\,,
\end{align} 
where $\kappa$ is the photon decay rate and $\Gamma$ is the spontaneous decay rate from $\ke{e}$ to $\ke{g}$.
The corresponding Lindblad operators are thus $a$ and $\sigmam_i$, respectively.
This choice of Lindblad operators is phenomenological and is adequate for the strong coupling regime~\cite{Scala2007, Betzholz2020}.

We now briefly discuss the eigenstates of the \gls{tc} Hamiltonian, shown in Eq.~\ref{eq:TCHamiltonian}, for excitation numbers beyond the well-studied case of one.
The Hamiltonian can be divided into blocks of the same cooperation number $S$~\cite{Campos-Gonzalez-Angulo2022-rz, Angulo_2021}.
These independent blocks are disconnected in the absence of dissipation and excitations of the molecules or the photon modes preserve $S$.

Figure~\ref{fig:ladder_2LS_N8}(a) shows the eigenvalues of Eq.~\ref{eq:TCHamiltonian} for $\omega_c=\omega_{eg}$ and $N=8$.
The vertical axis represents the excitation number ($\ev{\hat N_\text{x,TC}}$) and the horizontal axis represents the relative shift with respect to the uncoupled eigenvalues $\lambda$.
The inset numbers represent the number of degenerate states per eigenvalue, and the color code in (a) and (b) represents their normalized photon number ($\ev{\create\destroy}/\Nexc$) and in (c) and (d) the cooperation number $S$.
In Figs.~\ref{fig:ladder_2LS_N8}(b)-(d) the eigenstates are decomposed into different subgroups.
The ground state and the dark states, which have no photonic character and do not experience any splitting, are shown in Figure~\ref{fig:ladder_2LS_N8}(b). 
The dark states are only present for $\NexcTC\leq N/2$, due to their highly asymmetric nature, but their number grows quickly with $\NexcTC$~\cite{SupplementalMaterial,Andrilli_2010,Cline1976}.

\begin{figure*}
\centering
 \includegraphics[width=17cm]{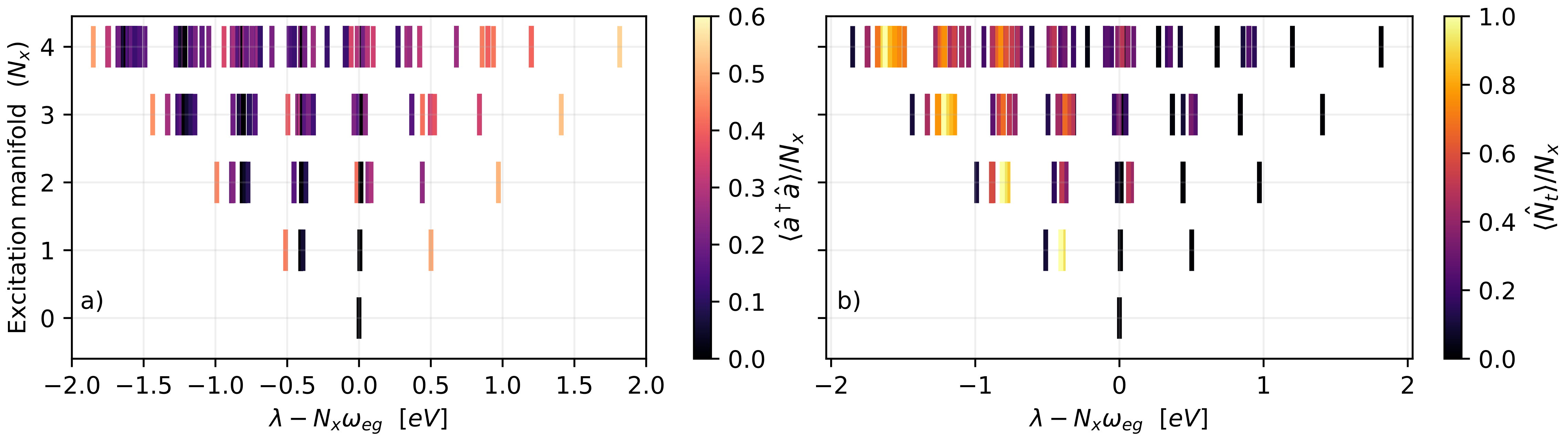} 
\caption{Energy diagrams for the eigenvalues ($\lambda$) of $N=8$ three-level molecules resonantly coupled to a single cavity mode. 
The eigenenergies are shown as excitation number versus relative, cavity induced shift.
The color code indicates the normalized expectation values of (a) $\create\destroy/\Nexc$ and (b) $\hat N_t/\Nexc$ for each eigenstate. The parameters are the same as in Fig.\ \ref{fig:ladder_2LS_N8} with additional parameter for the three-levelsystem: $c_{et} = 0.05\eV$, $\omega_{et}=0.4\eV $.}
\label{fig:ladder_3LS_N8}
\end{figure*}

Figure~\ref{fig:ladder_2LS_N8}(c) shows the polariton states ($\Nexc=1$) and the multi polariton states ($\Nexc>1$). 
The polariton and multi polariton states are nondegenerate and their number scales with $\Nexc+1$.
The $\Nexc$ excitations are distributed over all $N$ molecules, resulting in a small overall effect/contribution for an individual molecule~\cite{Scholes20jcpl,Davidsson2020-bs,Du2022-vj,Perez-Sanchez2023-mc}.
Nevertheless, these states are typically considered for chemical reactivity under strong coupling. 
Transitions within this group of states are possible in a "diagonal" direction via $\create,\destroy$ or $\Sigmap,\Sigmam$ and $\Delta S=0$.

Figure~\ref{fig:ladder_2LS_N8}(d) shows the dark polaritons grouped by the cooperation numbers $S < N/2$. 
These states are created by further excitation of the dark states with $\create$ or $\Sigmap$.
Their maximum energy splitting is smaller than the splitting of the multi polaritons~\cite{Ivanov, Ribeiro_2021, DelPo, Takemura,Campos-Gonzalez-Angulo2022-rz} (see Eq.~\ref{eq:Rabi}). 
However, these states appear in degenerate groups, and thus provide a combinatorial advantage over multi polaritons which can be populated during a photochemical reaction.
The ratio of dark polaritons to dark states for the dark polariton manifold generated by $\NexcTC-1$ can be estimated for relative excitation numbers $c=\NexcTC/N$~\cite{SupplementalMaterial,Andrilli_2010,Cline1976}:
\begin{align}\label{eq:darkstate_ratio}
    \dfrac{N_{DP}}{N_D} \approx \dfrac{\NexcTC}{N-\NexcTC}=\dfrac{c}{1-c}\,.
\end{align}
For $\NexcTC \gg 1$, this yields a significantly better ratio than for nondegenerate multi polaritons.

Next, we analyze the eigenvalues of the extended three-level model from Eq.~\ref{eq:3LS_Hamilt}.
Figure~\ref{fig:ladder_3LS_N8} shows the eigenvalues color-coded by (a) photon number and (b) $\ke{t}$ state excitation number.
The collective cavity coupling, $\sqrt{N}g_c$, is chosen so that the \gls{lp} state becomes resonant with $\ke{t}$.
The number of possible states is much larger and some of the degeneracies in Fig. \ref{fig:ladder_2LS_N8} are lifted.
However, some of the dark polariton group patterns, observed in Fig.~\ref{fig:ladder_2LS_N8}(a) can still be observed.
The upper multi polariton and dark polariton branch is largely unaffected by coupling with $\ke{t}$ and exhibits little to no $\ke{t}$ character, since the mixing is between $\ke{t}$ and the lower polariton branches.

\begin{figure}
    \centering
    \includegraphics[width=8.5cm]{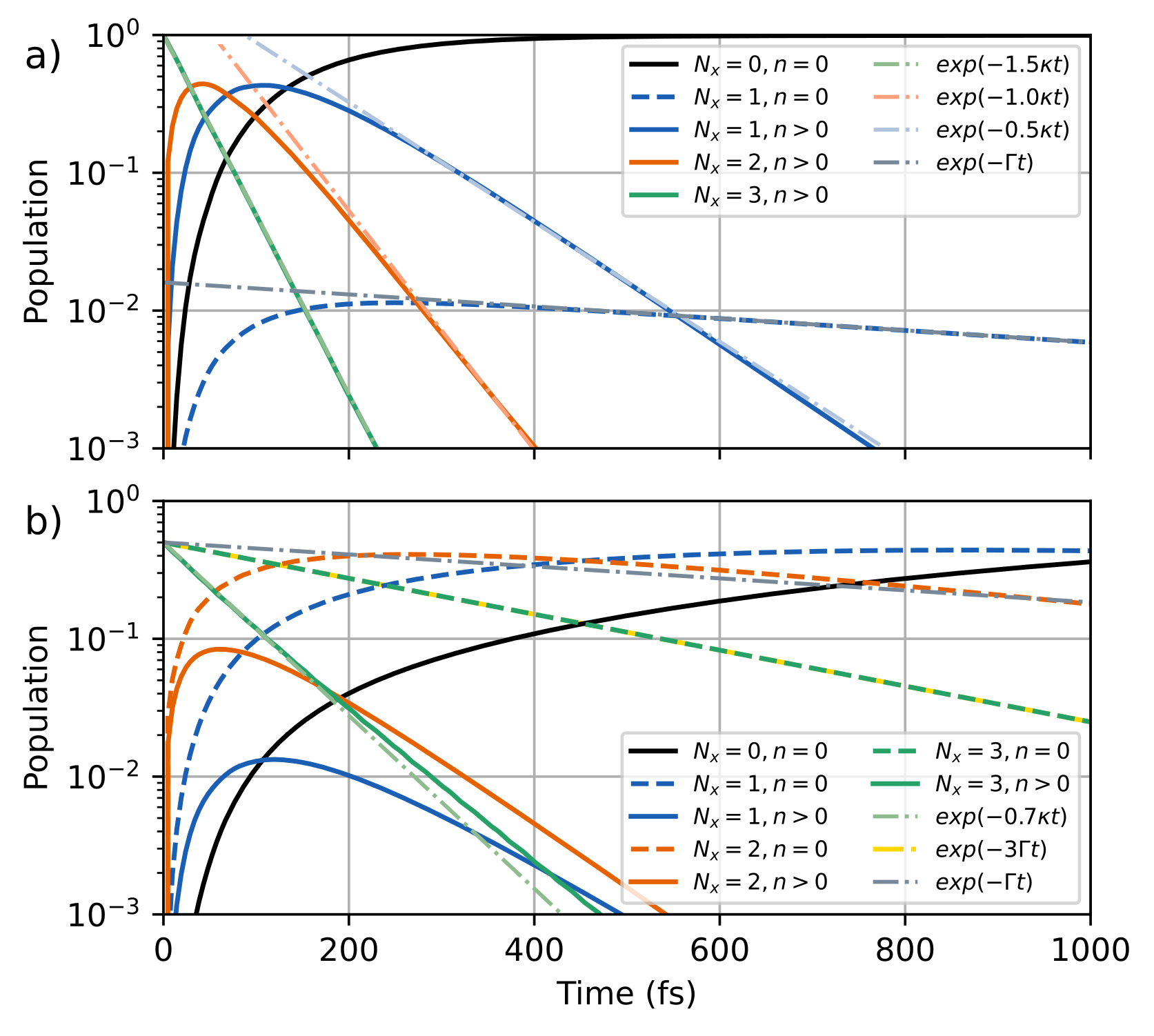}
    \caption{Population dynamics for $N=8$ two-level molecules 
    with (a) a initial state that represent a pure state with three molecular excitations $\Nexc=3$ and is a superposition of four multi polariton states
    and (b) an initial state that represents a mixed stats.
    The populations are grouped by the excitation number $\Nexc$ and photonic character.
    }    
    \label{fig:PropagationTLS_N8Ne3}
\end{figure}

In the following, we compare the dynamics of the \gls{tc} model with the two- and three-level systems in the presence of cavity and spontaneous decay.
The dynamics is calculated according to the master equation Eq.~\ref{eq:master_equation}.
The decay of the cavity photons is modeled as a fast process with $\kappa=0.02\,\fs^{-1}$ and the spontaneous decay of $\ke{e}$ as a slow process with $\Gamma=0.001\,\fs^{-1}$.
The cavity coupling between $\ke{g}$ and $\ke{e}$ is considered to be in the collective strong coupling regime with $\sqrt{N}g_c=0.5\,\eV$.
For details of the simulations, see the supplemental material~\cite{SupplementalMaterial,qutip,nix}.

Figure~\ref{fig:PropagationTLS_N8Ne3}(a) shows the time evolution of 8 two-level molecules, starting in a pure state with $\tr{\rho^2}=1$ and three excitations in $\ke{e}$.
This is the most ideal case, which represents a superradiant state.
The initial state density matrix has been constructed as $\rho = \out{\Phi}{\Phi}$, where $\ke{\Phi}=\sum_i\ke{\phi_i}$ is the sum of all bare states $\ke{\phi_i}$ with three excitations in $\ke{e}$,
and corresponds to a superposition of multi polariton states.
The initial state (green) decays fast on the time scale of cavity decay ($33$\,fs) and
passes rapidly through the intermediate excitation manifolds $\Nexc=2$ (orange) and $\Nexc=1$ (blue). 
Note that the effective cavity decay rate is scaled by the photon number.
At $\approx 600$\,fs nearly all population has reached the ground state (black line).
Spontaneous decay creates a minor amount ($10^{-2}$) of $\NexcTC=1$ dark states that decay as expected with rate $\Gamma$.

Figure~\ref{fig:PropagationTLS_N8Ne3}(b) shows the time evolution of 8 two-level molecules with their populations grouped into bright states ($n>0$) and dark states $n=0$ (solid and dashed lines, respectively).
The initial state is constructed similarly as for the pure state, 
but with minimal purity ($\tr{\rho^2} < 1$, $\rho_{i \neq j}=0$). 
Here, the density matrix of this mixed state is given by $\rho = \sum_i\out{\phi_i}{\phi_i}$, where $\ke{\phi_i}$ are all bare states with three excitations in $\ke{e}$.
This type of state could be expected when the molecular state $\ke{e}$ is populated by an incoherent excitation process or under thermal conditions via a chemical process.
Here, the initial state with $\NexcTC=3$ is composed of 50\% dark state character (green, dashed line)
and 50\% multi polariton and dark polariton character (green, solid line).
The dark state contribution decays according to the spontaneous decay rate ($\NexcTC\Gamma$, yellow dashed-dotted line).
The initial dark polariton and multi polariton states decay almost with the cavity decay rate ($0.7\kappa$, green dashed-dotted line).
The polarition states with $\NexcTC=2$ and $\NexcTC=1$ also decay with their scaled decay rates, respectively.
However, due to the contribution of different $S$ to the initial states, a large part decays into dark states with $\NexcTC=2$ (orange, dashed line) and further into $\NexcTC=1$ (blue, dashed line).
The dark states (dashed lines) can only decay by means of spontaneous decay. 

Subsequently, we show the dynamics for a corresponding three-level system with eight molecules. The parameters are identical to those of the two-level system.
The $\ke{t}$ state is 0.4\,eV below $\ke{e}$ and the coupling between $\ke{t}$ and $\ke{e}$ is $c_{et}=0.05$\,eV, which corresponds to a 83\,fs timescale and is in the perturbative regime. 
The initial states are now composed of 3 excitations in $\ke{t}$ instead of $\ke{e}$.

\begin{figure}
    \centering
    \includegraphics[width=8.5cm]{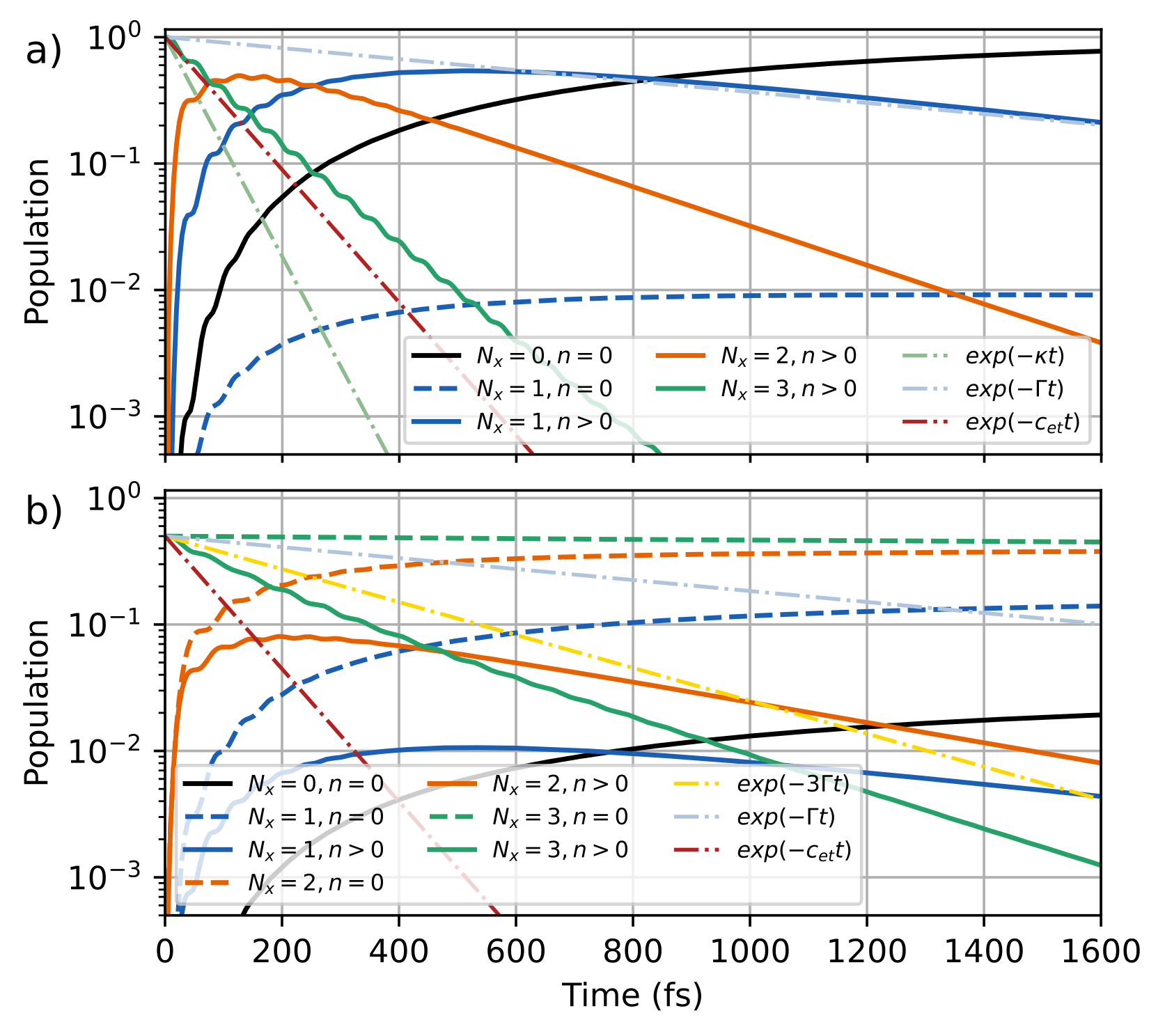}
    \caption{Dynamics of a three-level system with $N=8$ and an initial state 
    that represents (a) a pure and (b) a mixed state with three excitations in $\ke{t}$. The population is grouped by the excitation number, and photonic character.} 
    \label{fig:Propagation3LS_N8Nt3}
\end{figure}
Figure~\ref{fig:Propagation3LS_N8Nt3}(a) shows the time evolution starting in a pure state with $\tr{\rho}=1$ and three excited molecules in $\ke{t}$.
This state represents an ideal case where all molecules that are in state $\ke{t}$
have been prepared by a coherent process.
The initial state (green, solid line) 
thus is composed purely of multi polaritons from the
lower polariton branch with a varying amount of $\ke{t}$ character. 
This states decay rapidly, but its decay rate is now limited by the
coupling $c_{et}$ (visualized as a rate by the red dashed-dotted line).
The initial state decays almost exclusively into a mixture of polariton states with $\Nexc=2$ (orange, solid line), which in turn decays $\approx 3$ times slower than the initial state.
The subsequent populated states with $\Nexc=1$ also contain a small number of dark states (blue, dashed line), which are populated by spontaneous decay processes.
After 1\,ps $\approx$ 50\% has reached the ground state. The overall observed decay rates are now
slower than in the case of the two-level system
but are accelerated by the cavity compared to that of a system without a cavity~\cite{SupplementalMaterial}.

Figure~\ref{fig:Propagation3LS_N8Nt3}(b) shows the respective time evolution for the three-level system that starts in a mixed state.
Here, 50\% of the initial population is in dark states with a predominant $\ke{t}$ character
(green, dashed line), which shows an inhibited decay, since no direct decay channel for $\ke{t}$ is included.
The other 50\% of the initial population are composed of a mixture of dark polaritons and multi polaritons, with a varying amount of $\ke{t}$ character (green, solid line).
This state decays faster than the spontaneous decay (yellow, dashed-dotted line), but does not approach the full rate allowed by $c_{et}$.
Note that parts of the dark polariton states now decay in the $\Nexc=2$ dark states (orange, dashed line), where a large fraction of the population gets trapped.
In contrast, the $\Nexc=2$ polariton states (orange, solid line) decay further into $\Nexc=1$ polariton and dark states (blue, solid and dashed lines, respectively). At 1\,ps $\approx 98\%$ of the population is located in dark states, but no significant amount has reached the global ground state.
The overall decay of the $\ke{t}$ state is slower
than for the two-level system in Fig.~\ref{fig:PropagationTLS_N8Ne3}, but the decay
is again accelerated by coupling to the cavity compared to $\Nexc = 1$ (or without a cavity, see the supplemental material \cite{SupplementalMaterial}).

In Figs. \ref{fig:PropagationTLS_N8Ne3}-\ref{fig:Propagation3LS_N8Nt3}, we have shown a comparison of the ideal superradiant and a more realistic, incoherently prepared mixed state.
The superradiant state is a perfect superposition and represents a pure state in terms of the density matrix.
Such a state represents a linear combination of multi polariton eigenstates, or, in the case of the three-level system, a combination of multi polariton and dark polariton states. 
This state can reach the ground state very efficiently via cavity decay and is only limited by the cavity decay rate and, in the three-level model, additionally by the coupling between $\ke{t}$ and $\ke{e}$.
Low photon numbers, which can be a result of quantum interference between upper and lower polariton states, can also lead to a delayed decay of the excitation.
The mixed state, in contrast, could be expected for non-zero temperature conditions and represents an incoherent state preparation, and is therefore closest to realistic experimental conditions at room temperature.
This state projects on all possible eigenstates, namely the multi polariton states, the dark polaritons, and the dark states.
As can be seen in Fig.~\ref{fig:ladder_2LS_N8}, the number of degenerate states for each value of $S$ becomes important. Here, $2S$ represents the number of molecules that are effectively coupled to the cavity mode.
The multi polaritons with $S=N/2$ are not degenerate and thus provide the smallest number of states.
These states suffer from the $1/N$ problem~\cite{Scholes20jcpl,Du2022-vj,Perez-Sanchez2023-mc} and have a negligible contribution to the initial state.
For $\Nexc=1$ there are $N-1$ dark states that will dominate the superposition.
However, for $\Nexc >1$ dark polaritons are formed from dark states with $\Nexc-1$, $\Nexc-2$, ... and are present in a much larger number. 
In the discussed examples of two-level systems with $\NexcTC=3$, the ratio of multi polariton and dark polariton states to dark states is approximately 2:1.
This ratio improves with smaller cooperation numbers $2S$ (see Eq.~\ref{eq:darkstate_ratio}), providing more states for hybridization and reactivity. 
However, dark polaritons introduce a reduced overall collective coupling strength, scaling with $\sqrt{2S}$, because the excitations are distributed over fewer molecules.
Note that the observed Rabi-splitting depends on $S$, but is otherwise nearly independent of the total number of excitations.

The generalization to large $N$ (Eq. \ref{eq:darkstate_ratio}) shows that relative number
of excited molecules is decisive.
For example, for $S=N/2-\Nexc+1$ (i.e. the dark polaritons generated by exciting $\Nexc-1$) and a relative excitation number $c=0.01$ the ratio of dark polariton states to dark states is $1:100$, and thus there are sufficiently many bright states available for interaction with the cavity mode.
The Rabi splitting in this scenario is still 99\% of the maximum possible Rabi splitting, leaving the effective cavity coupling nearly unaffected.
Together with the observation in Fig.~\ref{fig:Propagation3LS_N8Nt3}(b), this suggests that polaritonic chemistry reactions may require a fraction of the sample to be excited to make use the collective coupling.

Dark polaritons cannot reach the global ground state because they decay to dark states with the same $S$, which limits the number of molecules that can undergo a cavity-assisted reaction.
The overall decay is still more efficient than in the absence of cavity decay or in the absence of strong coupling.
As shown in Fig.~\ref{fig:Propagation3LS_N8Nt3}(b), only a fraction of molecules experiences accelerated decay, but the overall efficiency increases strongly for $\Nexc > 1$ compared to $\Nexc = 1$ (see the supplemental material \cite{SupplementalMaterial}). 

It is important to note that linear spectroscopy alone may not be able to distinguish the state of the system in terms of the excitation number. 
Optical transitions are allowed only for $\Delta \Nexc = \pm1$ and $\Delta S=0$.
The observed Rabi-splitting scales with $\sqrt{2S/V_c}$, and thus the density of molecules that are effectively coupled to the cavity modes. 
The total number of excitations $\Nexc$ has no significant effect on the observed Rabi splitting, in the limit of large $N$ and $\Nexc \ll N$.
More sophisticated spectroscopic schemes are thus needed to distinguish these additional degrees of freedom.

In summary, we have shown that states with higher excitation numbers in an ensemble of molecules resonantly coupled to a cavity provide a favorable density of states.
The investigated three-level molecular model has broad applicability and shows that a photochemical reaction is still possible, despite the presence of a large number of dark states.
Shifting the focus away from the (multi) polariton states, it is the group of dark polaritons that provides a large number of states to interact with, and thus gives rise to an entropic advantage.
These states allow to overcome the $1/N$ problem in polaritonic chemistry which is typically cited as a paradox~\cite{Scholes20jcpl}.
The introduced model represents, for example, a triplet-triplet annihilation process~\cite{Ye2021-sc}, where a significant fraction of molecules are created in an excited state and the reaction is driven by photon decay.
Our analysis also hints that in electronic strong coupling in general,
only a fraction of the molecules in the cavity take part in a reaction.
Moreover, higher excitation numbers may be needed to enable polaritonic reactions.
It thus can be expected that the states that appear in the
higher excitation manifolds are important for a statistical
mechanics model that aims to describe reaction rates.
We have analyzed ensembles with a small number of molecules in the absence of energetic disorder and nuclear degrees of freedom.
Taking into account these types of disorder present at finite temperature introduces mixing~\cite{Davidsson2023-xa,Wellnitz21jcp,Wellnitz2022-xb,Zhou2023-zo,Mattiotti2024-un} of polariton states and dark states and thus could be expected to further modify the density of states.

Our findings may have broader implications for the understanding of polaritonic chemistry as it sheds light on the underlying statistics and mechanism.
The presented model fits well for electronic strong coupling, which involves electronically excited states. However, without loss of generality this model could be extended to investigate the statistics of vibrational strong coupled systems.

\section*{Acknowledgment}
The authors thank Johannes Schachenmayer for helpful feedback on the manuscript. This project has received funding from the European Research Council (ERC) under the European Union’s Horizon 2020 research and innovation program (grant agreement no. 852286). Support from the Swedish Research Council (Grant No.~VR 2024-04299) is acknowledged.

\section*{Author Contributions}
L.B. conducted the investigation; T.S. and M.K. conceptualized the work. All authors contributed equally to writing the draft of the manuscript.

\section*{Data Availability}
The data supporting this study's findings are available from the corresponding author upon reasonable request.

\bibliography{lit.bib}

\end{document}


\title{Supporting Information: \\The role of dark polariton states for electronic strong coupling in molecules}

\author{Lucas Borges}
\affiliation{Department of Physics, Stockholm University, AlbaNova University Center, SE-106 91 Stockholm, Sweden}

\author{Thomas Schnappinger}
\affiliation{Department of Physics, Stockholm University, AlbaNova University Center, SE-106 91 Stockholm, Sweden}

\author{Markus Kowalewski}
\email{markus.kowalewski@fysik.su.se}
\affiliation{Department of Physics, Stockholm University, AlbaNova University Center, SE-106 91 Stockholm, Sweden}

\maketitle

\tableofcontents
\clearpage

\section{Scaling of dark state number in higher excitation manifolds}

A state $\ke \psids$ formed by $N$ two-level molecules strongly coupled to a cavity mode is considered a dark state if it is an eigenstate of the Hamiltonian $\hamilt$ and satisfies $\expcv{\psids}{\create\destroy}{\psids} = 0$.
For a manifold with $\NexcTC$ excitations, 
the number of states that involve only excitations in the matter part and thus have zero photonic character
is given by the binomial coefficient
$\M = \binom{N}{\NexcTC}$. 
Therefore, a dark state $\ke{\psids}$ within this manifold must be a linear combination of these $\M$ bare states, which can be expressed as:
\begin{equation} \label{eq:psiDS}
    \ke{\psids} = \sum_i^\M \alpha_i \mS^{\NexcTC}_i\ke{0},
\end{equation}
with 
\begin{equation}
\mS^{\NexcTC}_i = \bigotimes_{k\in \mathcal{C}_i} \sigmap_{k},
\end{equation}
where $\mathcal{C}_i$ is the \ith set of the combinations of $\{1, \dots, N\}$ choose $\NexcTC$.
The action of $\mS^{\NexcTC}_i$ on the collective ground state $\ke{0}$ produces the \ith bare state, with $\alpha_i$ being its contribution to the dark state in Eq.~\eqref{eq:psiDS}.

The action of the Hamiltonian on $\ke{\psids}$ is given by
\begin{align}
\hamilt_{TC}\ke{\psids} &= \sum_{k=1}^N\left(\omega_{eg}\ke{e_k}\br{e_k}  + g_c \create\sigmam_k \right)\ke{\psids} ,\\
    \hamilt_{TC}\ke{\psids} &= {\NexcTC}\omega_{eg}\ke{\psids} + g_c \sum_{j=1}^{\M'} \beta_j \mS_j^{{\NexcTC}-1}\create\ke{0},
\end{align}
where $\M' = \binom{N}{\NexcTC}$, and $\beta_j$ are sums of the coefficients $\alpha_i$ such that $\mS^{{\NexcTC}-1}_j = \sigmam_k\mS^{\NexcTC}_i$, for any $k$.

For $\ke{\psids}$ to be an eigenstate of $\hamilt$, requires that $\beta_j = 0, \forall j$, which is equivalent to a homogeneous system of $\M'$ linear equations with $\M$ parameters. 
Infinite nontrivial solutions for this system exist only when $\M > \M'$~\cite{Andrilli_2010,Cline1976}.
The total number of dark states in the manifold is then the dimension of the non-trivial solutions for the homogeouns system of linear equations $\beta_j=0$, which corresponds to the difference between $\M$ and $\M'$, and reads:
\begin{equation}
    \Nds(N, \NexcTC) = \binom{N}{\NexcTC} - \binom{N}{\NexcTC-1} = \frac{N!(N-2\NexcTC+1)}{\NexcTC!(N-\NexcTC+1)!}.
    \label{eq:Nds}
\end{equation} 
Their cooperation number is given by $S=N/2-\NexcTC$.
Consequently, we can deduce that there are no dark states for manifolds with $\NexcTC > N/2+1$. 

\begin{figure}
\includegraphics[width=0.8\textwidth]{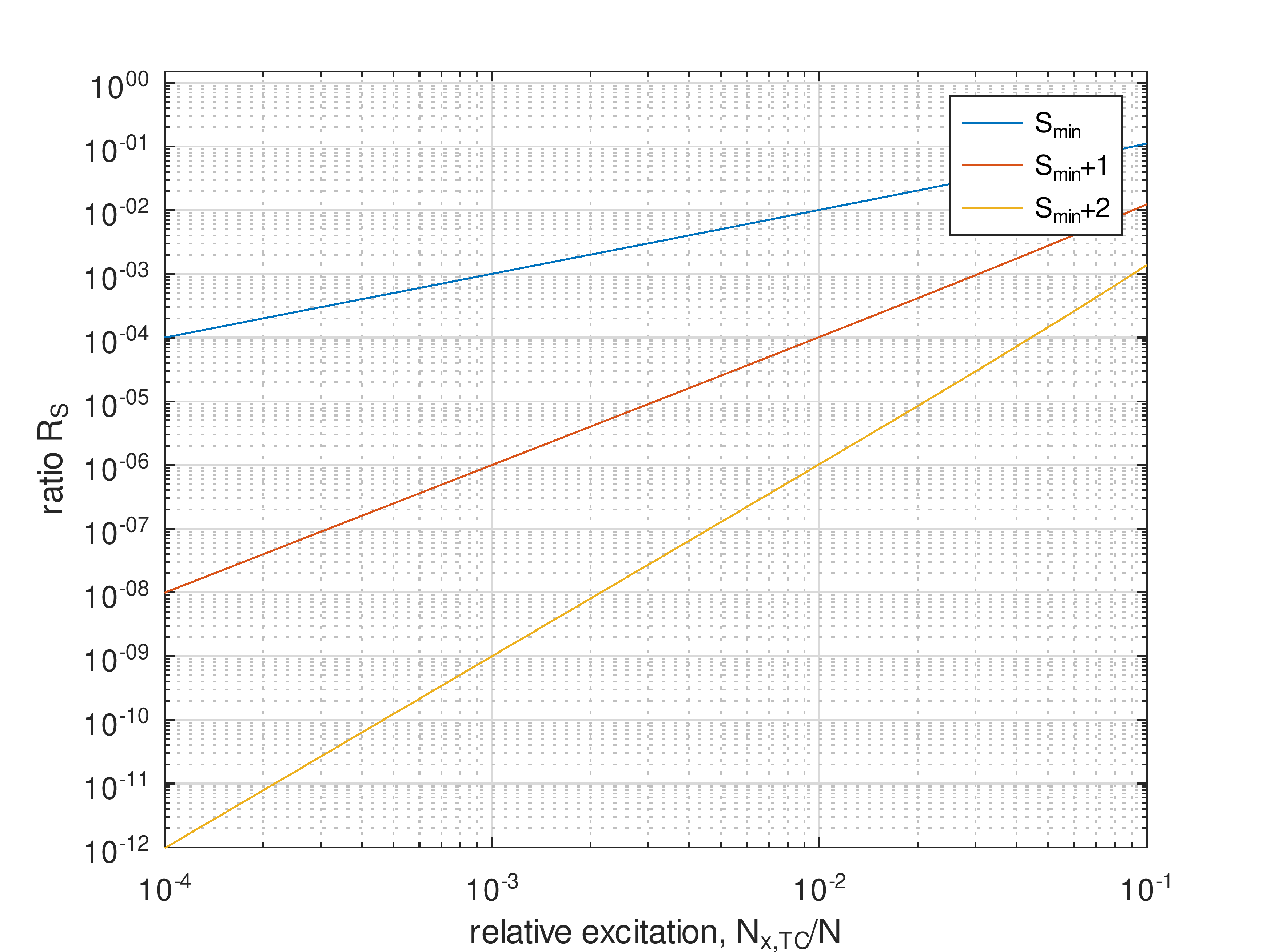}
\caption{Dark polariton to dark state ratios in dependence of the
relative excitation number, according to Eq.~\ref{eq:ratio} for $S=N/2-\NexcTC+1$ (blue), $S=N/2-\NexcTC+2$ (red), and $S=N/2-\NexcTC+3$.}
\label{fig:ratio}
\end{figure}
\begin{figure}
\includegraphics[width=0.8\textwidth]{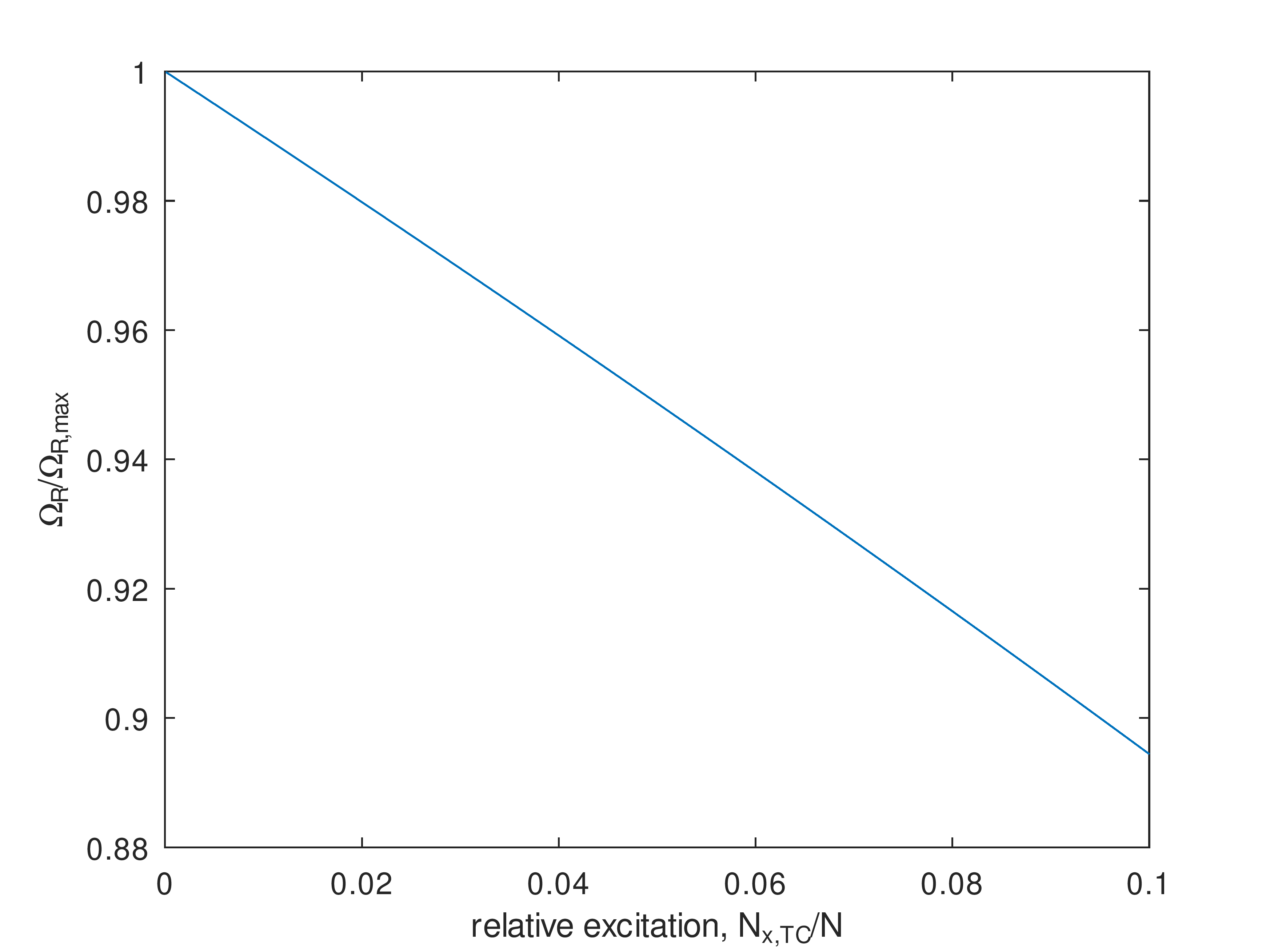}
\caption{Rabi splitting relative to the maximum possible Rabi splitting according to Eq.~\ref{eq:Omega_rel}.}
\label{fig:rabi_rel}
\end{figure}

For the case of the second excitation manifold, the number of dark states is $\Nds(N, 2) = N(N-3)/2$, while the number of dark polaritons is $2(N-1)$, arising from the splitting of the dark states present in the single-excitation manifold.
The ratio of the lower dark polariton states to dark states can be estimated using Eq.~\ref{eq:Nds} and the fact that dark polaritons are created by further exciting dark states.
For dark polaritons with a given $\NexcTC$ and a given quantum number $S$ the ratio can be written as:
\begin{align}
R_S=\dfrac{\Nds(N, \NexcTC/2 -S)}{\Nds(N, \NexcTC )} =  
  \dfrac{(2S+1)\prod_{p=N/2-S+1}^{\NexcTC} p}{(N-2\NexcTC +1)\prod_{p=N-\NexcTC+2}^{N/2+S+1} p}\,. \label{eq:ratio}
\end{align}
The minimum $S$ for a dark polariton manifold with given $\NexcTC$ is given
by $S_i=N/2-\NexcTC+i$ with $i=1$.
In the limit of large $N$ and $S \ll N/2$ and small $i$, Eq.~\ref{eq:ratio} can approximated as
\begin{align}
    R_{S} = \dfrac{\Nds(N, \NexcTC/2 -S_i)}{\Nds(N, \NexcTC )} \approx  
  \dfrac{\prod_{p=N/2-S_i+1}^{\NexcTC} p}{\prod_{p=N-\NexcTC+2}^{N/2+S_i+1} p} \approx \left(\dfrac{\NexcTC}{N-\NexcTC}\right)^i
  = \left(\dfrac{c}{1-c}\right)^i\,,
\end{align}
where we have replaced the excitation number with the relative excitation number $c=\NexcTC/N$.
Figure~\ref{fig:ratio} shows the ratios in dependence of $c$.
The Rabi splitting relative to the maximum possible value is given by $\Omega_{rel}=\sqrt{2S_i/N}$, which for large $N$ yields: 
\begin{align}
\Omega_{rel}\approx \sqrt{1-\dfrac{2\NexcTC}{N}}=\sqrt{1-2c}
\,.
\label{eq:Omega_rel}
\end{align}
Note, that for large $N$ and small $i$, $\Omega_{rel}$ does not depend on $i$.
Figure \ref{fig:rabi_rel} shows the relative Rabi splitting in dependence of $c$.

\clearpage
\section{Simulation details}

The \gls{tc} Hamiltonian for two-level and three-level systems coupled to a single cavity mode, shown in Eqs.~(1) and~(6) in the manuscript, was constructed and numerically diagonalized using the QuTiP package~\cite{qutip}, version 5.0.2. 
The open system dynamics described by Eq.~(8) in the manuscript, is solved using the QuTiP package mesolve, which is based on the SciPy zvode ODE solver.
All calculations were performed in a reproducible environment using the Nix package manager together with NixOS-QChem~\cite{Kowalewski2022-zt} (commit 319ff67).

We considered systems with $N=8$ molecules and a maximum of $N_x=4$ excitations located in the molecular subsystem.
The initial state for the dynamics is constructed with $N_x=3$ excitations located in the molecular subsystem, and the Hamiltonian is truncated at the corresponding third excitations manifold.
The open system dynamics is performed using adaptive time steps and the state populations are calculated every 5\,fs. 
The decay rates for the dynamics are $\kappa=1/50$\,fs$^{-1}$ and $\Gamma=1/1000$\,fs$^{-1}$.
To compare the dynamics of ensembles with different numbers of molecules, we maintain a constant collective coupling strength $g_c\sqrt{N}$ by scaling the single molecule coupling strength to ensure a consistent Rabi splitting for ensembles of different sizes.

The uncoupled basis for \gls{tc} Hamiltonian can be written as products of the bare molecular states $\ke{\phi_i} \in \ke{g_i},\ke{e_i}$ and the Fock states $\ke{n}$ of the cavity mode:
\begin{gather}\ke{\Psi}=\ke{\phi_1}\otimes\cdots\otimes\ke{\phi_N}\otimes\ke{n}\,.\label{eq:product-states}
\end{gather}
The total number of states for a given number of excitations $\NexcTC \le N$ is given by the sum of the binomial coefficients $\sum_{i=0}^{\NexcTC}\binom{N}{i}$, corresponding to all combinations of $\NexcTC$ excitations among the $N$ molecules and the excitations of the cavity mode. 
In this study, we only take into account the cases where $\NexcTC \le N$.
In the absence of counter-rotating coupling terms, the ground state of the coupled system with $\NexcTC=0$ is simply given by the product of all molecular ground states and the ground state of the cavity mode: $\ke{G,0} = \bigotimes_{i=1}^N \ke{g_i}\otimes \ke{n=0}$.

\clearpage

\section{Supplemental plots for the dynamics}

\begin{figure}
    \centering
    \includegraphics[width=0.8\linewidth]{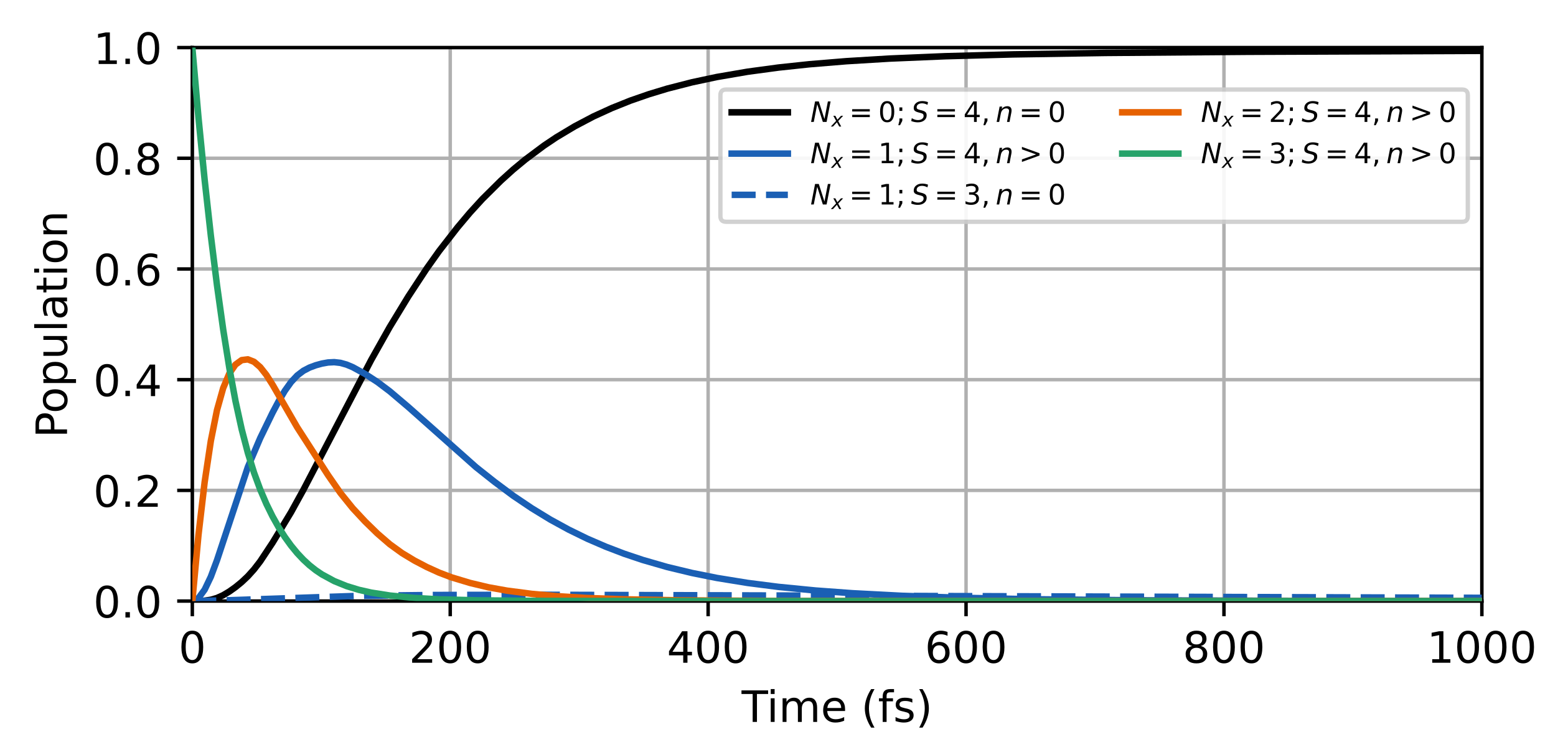}
    \caption{   
Population dynamics for $N = 8$ two-level molecules with a initial state that represent a pure state with three molecular excitations and is a superposition of four multi polariton states with $\NexcTC = 3$ and $S = 4$. 
The populations are grouped by the excitation number $\NexcTC$, cooperation number $S$ and photonic character. 
Only polariton states with $S = 4$ contribute to the populations.
}
    \label{fig:TLS_pure}
\end{figure}

In this section, we provide additional plots for the dynamics of the two-level and three-level systems discussed in the manuscript. 
Fig.~\ref{fig:TLS_pure} shows the time evolution of eight two-level molecules on a linear scale, starting in the pure state with $\tr{\rho^2}=1$ and three excitations in $\ke{e}$.
The parameters used are $g_c=0.5\eV/\sqrt{N}$, $\omega_c=\omega_{eg}=4,3\,\eV$, $\kappa=1/50\,\fs^{-1}$, and $\Gamma=1/1000\,\fs^{-1}$.
The populations are grouped by the number of excitations, the cooperation number, and the photonic character, showing that the dynamics involves mostly multi polaritons states with $S=N/2$. Only a small contribution from dark states is observed (blue dashed curve).

\begin{figure}
    \centering
    \includegraphics[width=0.8\linewidth]{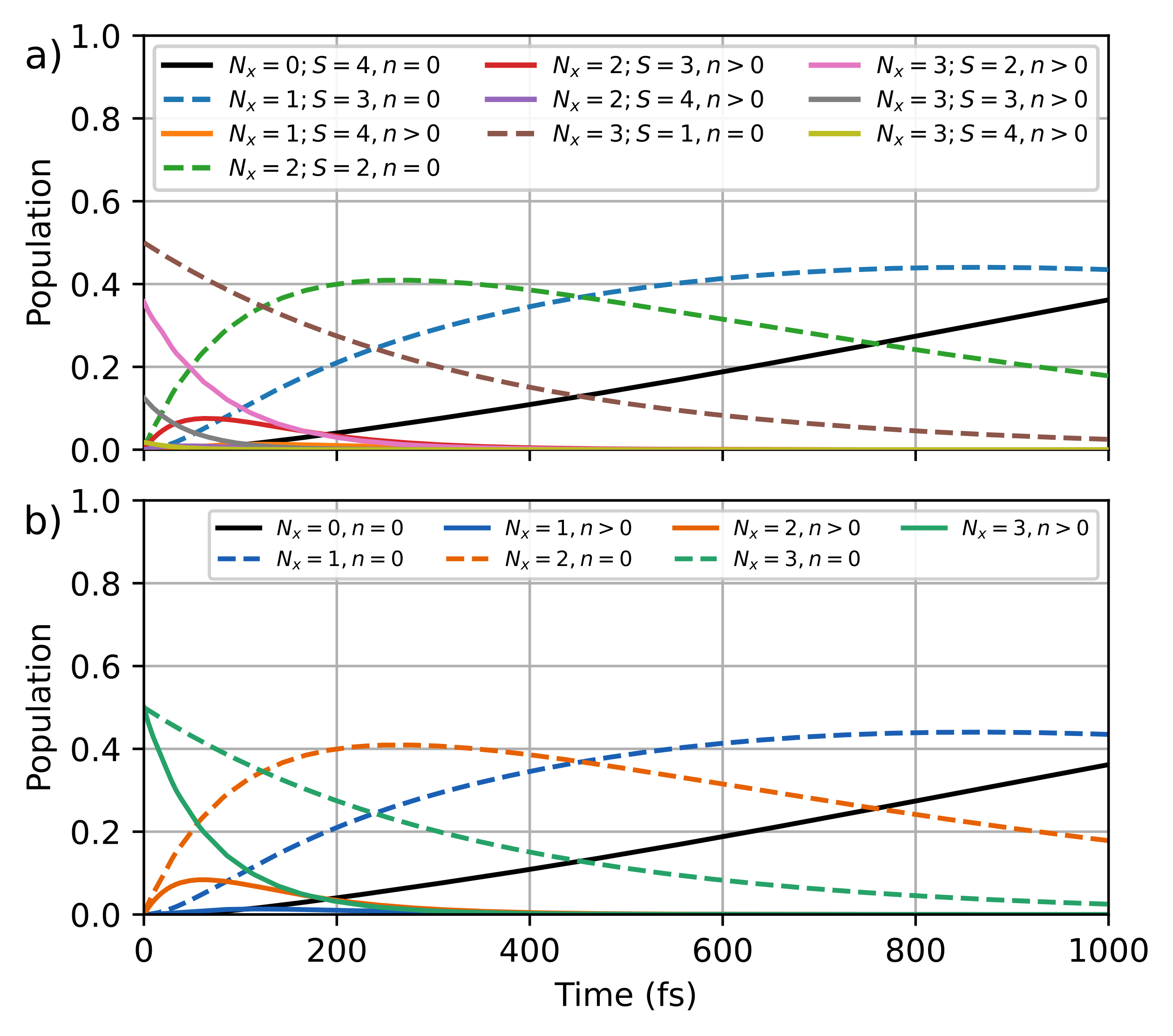}
    \caption{Population dynamics for $N = 8$ two-level molecules with a initial state that represent a mixed state with three molecular excitations and is a superposition of multi polaritons , dark polaritons and dark states with $\NexcTC = 3$. The populations are grouped in (a) by the excitation number $\NexcTC$, cooperation number $S$ and photonic character and in (b) only by $\NexcTC$ and photonic character.}
    \label{fig:TLS_impure}
\end{figure}

The time evolution of 8 two-level molecules is shown in Fig.~\ref{fig:TLS_impure} on a linear scale, starting in a mixed state with $\tr{\rho^2}<1$ and three excitations in $\ke{e}$.
In Fig.~\ref{fig:TLS_impure}(a) the populations are grouped by the excitation number $\NexcTC$, cooperation number $S$ and photonic character.
The initial mixed state consists of dark states ($50\%)$, dark polaritons ($35,7\%$ and $12,5\%$), and multi polaritons ($1,8\%$).
The states shown in Fig.~\ref{fig:TLS_impure}(b) are grouped only by the excitation number $\NexcTC$ and photonic character.

\begin{figure}
    \centering
    \includegraphics[width=0.8\linewidth]{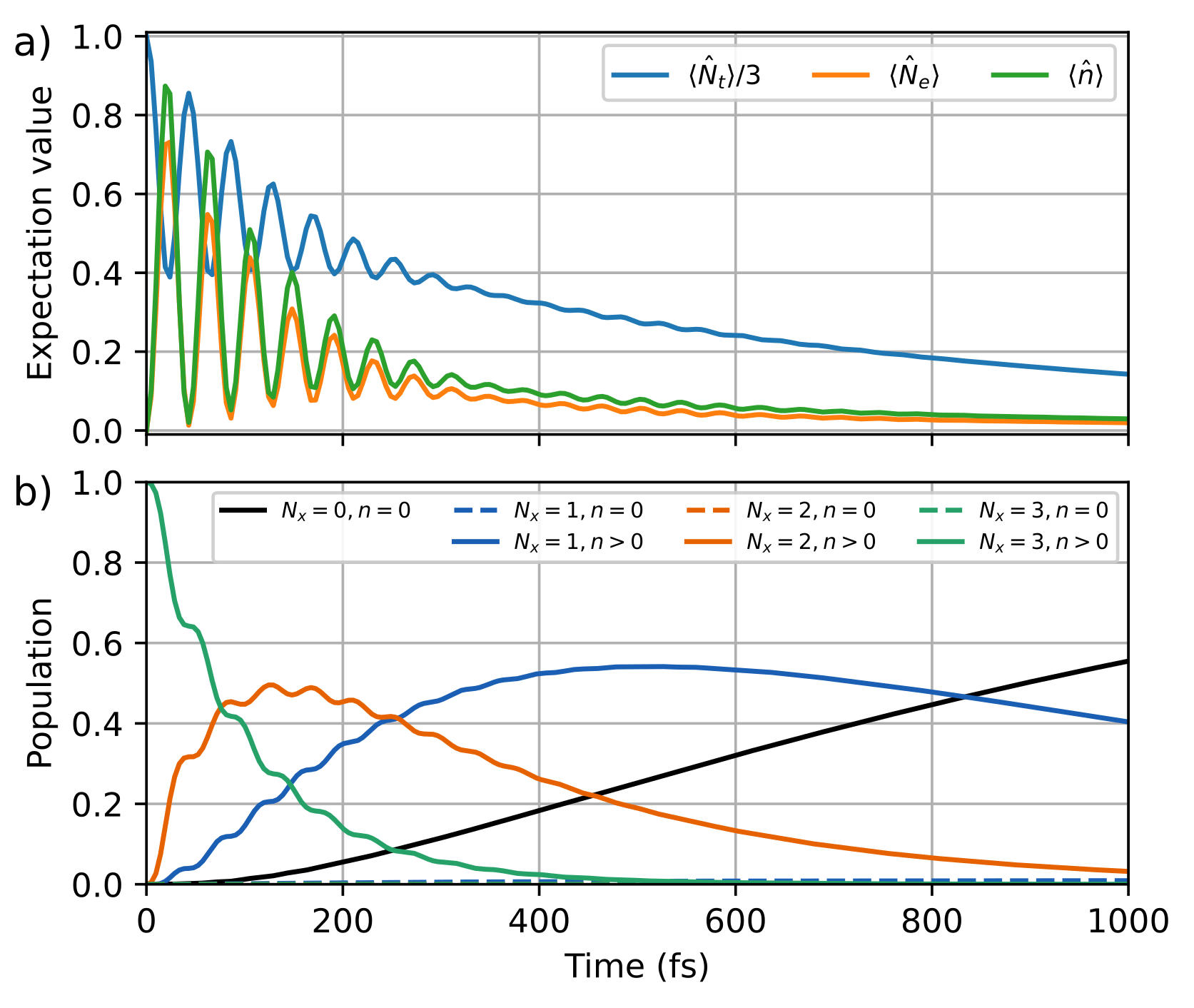}
    \caption{
    Dynamics of a three-level system with $N = 8$ and an initial state that represents a pure state with three excitations in $\ke{t}$. 
(a) Expectation values of $\hat N_t$, $\hat N_e$ and $\hat n$, giving the quantity of $t$, $e$ and photonic excitations in the system.
(b) Population grouped by the excitation number and photonic character.
}
    \label{fig:3LS_pure}
\end{figure}

\begin{figure}
    \centering
    \includegraphics[width=0.8\linewidth]{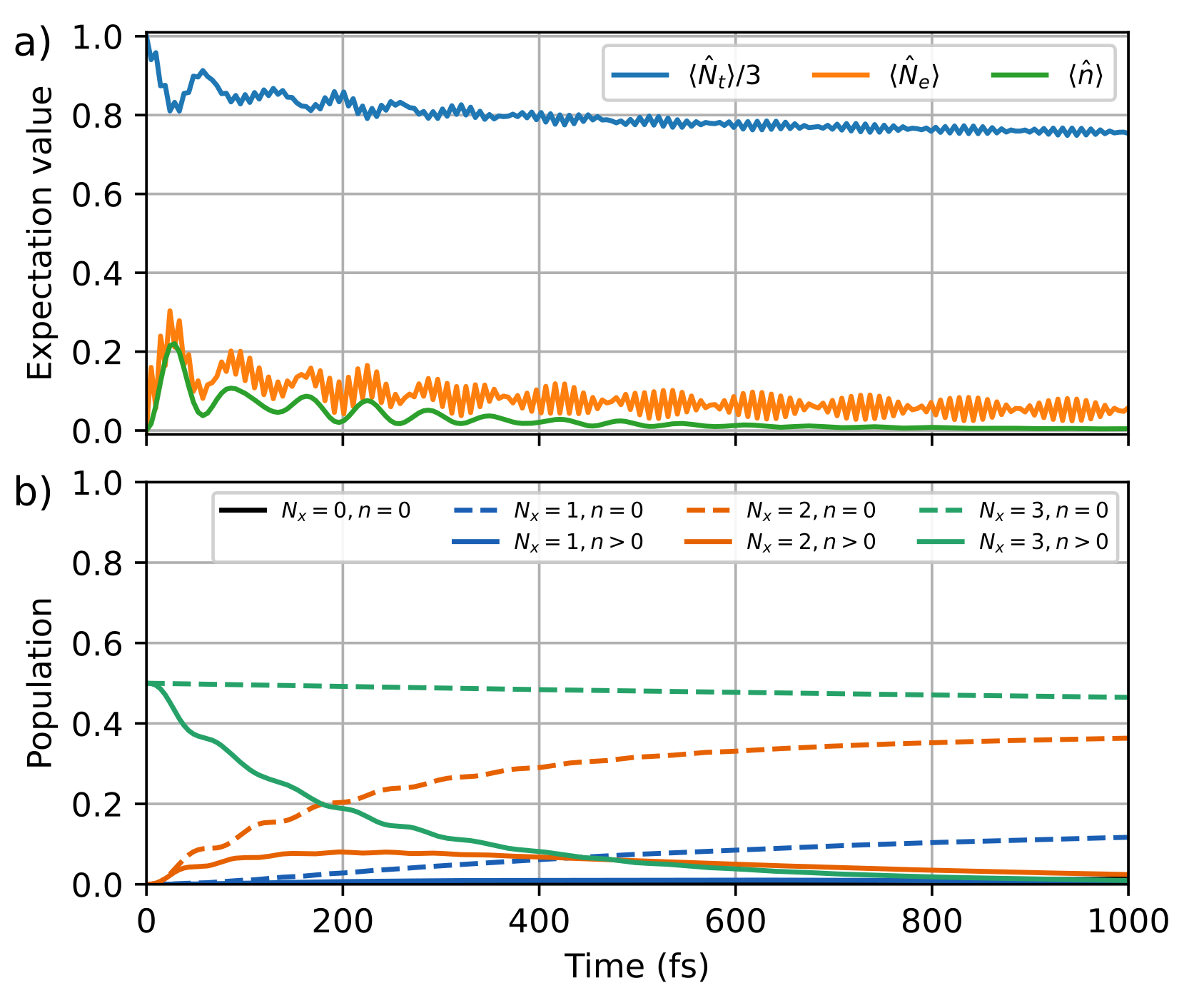}
    \caption{
    Dynamics of a three-level system with $N = 8$ and an initial state that represents a mixed state with three excitations in $\ke{t}$. 
(a) Expectation values of $\hat N_t$, $\hat N_e$ and $\hat n$, giving the quantity of $t$, $e$ and photonic excitations in the system.
(b) Population grouped by the excitation number and photonic character.
}
    \label{fig:3LS_impure}
\end{figure}

In Figs.~\ref{fig:3LS_pure} and \ref{fig:3LS_impure} we show the dynamics of three-level systems with $N = 8$ and initially three excitations in $\ke{t}$ on a linear scale. 
The parameters used are identical to those of the two-level system. The $\ke{t}$ state is 0.4\,eV below $\ke{e}$ and the coupling between $\ke{t}$ and $\ke{e}$ is $c_{et}=0.05$\,eV.
The initial state for the dynamics shown in Fig.~\ref{fig:3LS_pure} has maximum purity, $\tr{\rho^2}=1$, and for Fig.~\ref{fig:3LS_impure} it has minimum purity.
The expectation values of the operators $\hat N_t$, $\hat N_e$ and $\hat n$ are shown in Fig.~\ref{fig:3LS_pure}(a) for the pure initial state and in Fig.~\ref{fig:3LS_impure}(a) for the mixed initial state. 
In the case of a pure initial state, we observe significant oscillations between $N_t$ and a combination of $N_e$ and $n$, indicating an effective exchange between $\ke{t}$ and the cavity photon via $\ke{e}$. 
As a consequence, a sufficiently large number of photons allows an efficient decay channel through the cavity to the global ground state, leading to a fast decrease of all three expectation values over time.
In contrast, for the mixed initial state, the three expectation values behave differently.
Within the first hundred femtoseconds, $N_t$ decays three times slower than in the pure case, and for later times the decay slows down even more. 
As a result, $N_e$ and $n$ remain very small, and a fast decay via the cavity to the global ground state is less efficient than in the case of pure initial state.
In Fig.~\ref{fig:3LS_pure}(b) and Fig.~\ref{fig:3LS_impure}(b) the state population is shown on a linear scale grouped by number of excitations and photonic character for both pure and mixed initial states.

\begin{figure}
    \centering
    \includegraphics[width=0.8\linewidth]{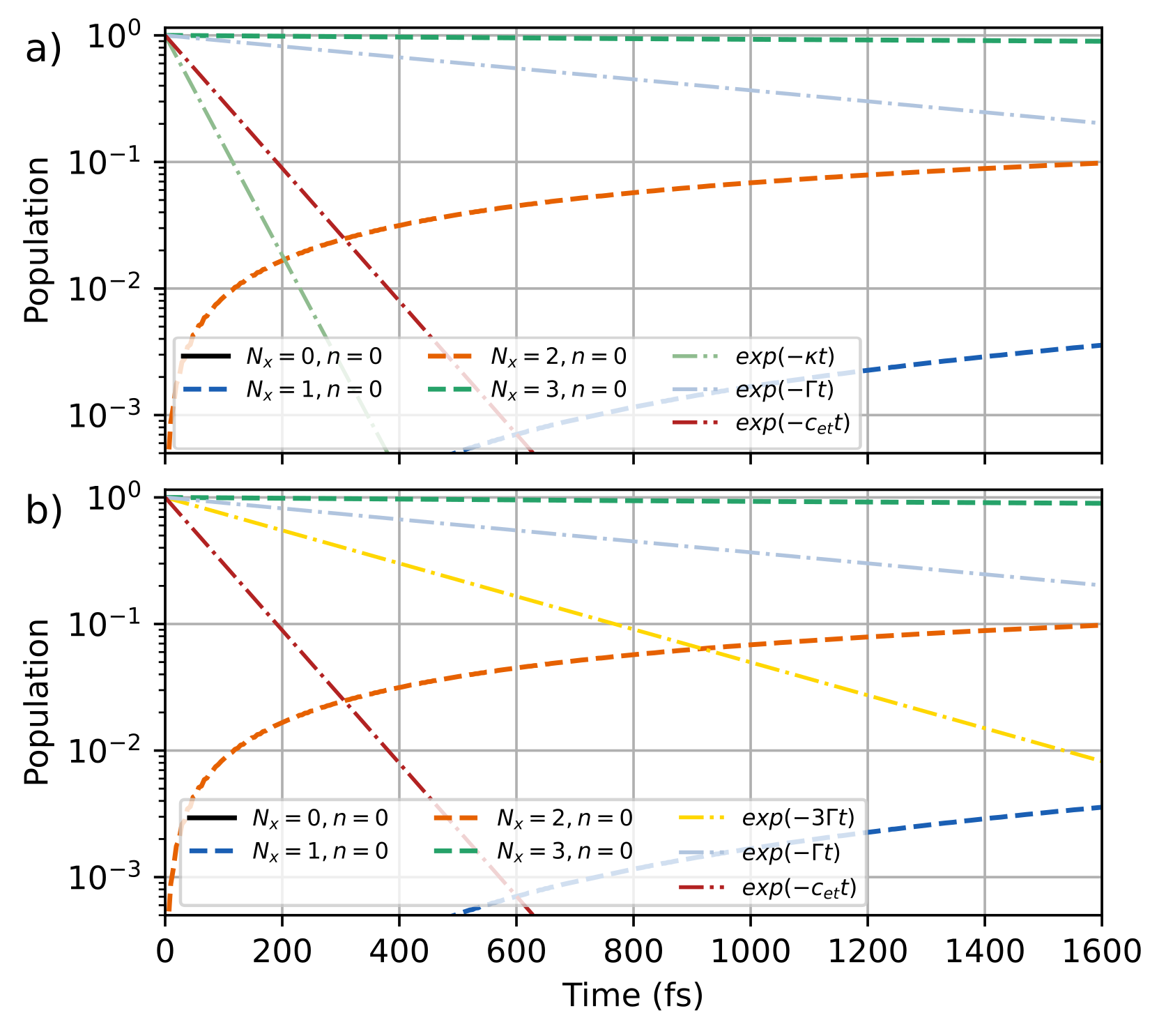}
    \caption{
    Dynamics of a three-level system with $N = 8$ and an initial state that represents a (a) pure and (b) mixed state with three excitations in $\ke{t}$, without cavity coupling between levels $e$ and $g$ ($g_c=0$\,eV).
The population is grouped by the excitation number.}
    \label{fig:nocavity}
\end{figure}

As a reference for the three-level system with $N = 8$, we show in Fig.~\ref{fig:nocavity} the evolution of the system with initially three excitations in $\ke{t}$ outside a cavity by setting the cavity coupling to zero.
Whether the initial state is pure, shown in Fig.~\ref{fig:nocavity}(a), or a mixed state, shown in Fig.~\ref{fig:nocavity}(b), the observed population dynamics are very slow and nearly identical. 
The only way to reach lower excitation manifolds is the spontaneous decay from the $\ke{e}$ state, which is slowly populated since $\ke{e}$ and $\ke{t}$ are detuned without coupling to the cavity.

Additionally, we show the dynamics of the three-level system with $N = 8$ resonantly coupled to a cavity mode and starting in a mixed state with three excitations in $\ke{t}$ in Fig.~\ref{fig:cets} for different $c_{et}$ couplings.  
All other parameters are identical to Fig.~\label{fig:cets}. 
For smaller coupling, $c_{et}=0.025\,\eV$ (half the initial value), shown in Fig.~\ref{fig:cets}(a), the overall decay slows down significantly, while for large coupling, $c_{et}=0.1\,\eV$ (double the initial value), shown in Fig.~\ref{fig:cets}(b), the decay in the global ground state increases.
\begin{figure}
    \centering
    \includegraphics[width=0.8\linewidth]{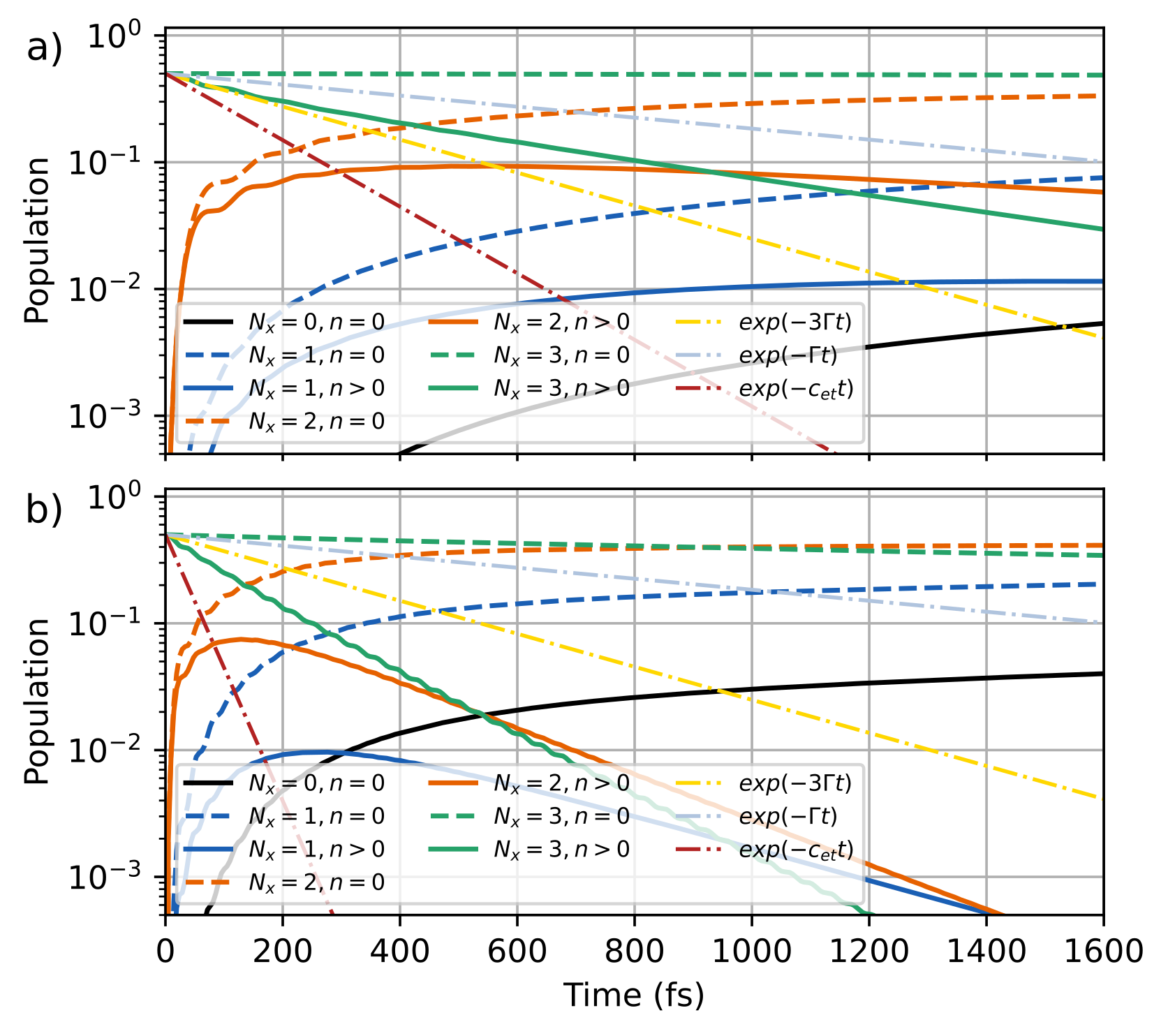}
    \caption{
Dynamics of a three-level system with $N = 8$ and an initial state that represents a mixed state with three excitations in $\ke{t}$, with coupling between levels $e$ and $t$ given by (a) $c_{et}=0.025$\,eV and (b) $c_{et}=0.1$\,eV.
The population is grouped by the excitation number and photonic character.
}
    \label{fig:cets}
\end{figure}
\clearpage

\bibliography{lit.bib}